\documentclass[aps,prl,10pt,twocolumn,footnoteinbib,superscriptaddress]{revtex4-1}
\def\arXiv{1}  % Is this manuscript for an arXiv preprint?

\usepackage{graphicx}
\usepackage{here}
\usepackage{amsbsy,amssymb,amsmath,bm,mathtools}

\usepackage{times,txfonts}

\DeclareFontFamily{OT1}{pzc}{}
\DeclareFontShape{OT1}{pzc}{m}{it}{<-> s * [1.2] pzcmi7t}{}
\DeclareMathAlphabet{\mathpzc}{OT1}{pzc}{m}{it}

\usepackage{xspace}
\usepackage{hyperref}
\usepackage[usenames]{color}
\usepackage[dvipsnames]{xcolor}
\hypersetup{
  colorlinks   = true,
  citecolor    = blue,
  linkcolor    = blue,
  urlcolor     = blue
}

\newcommand{\ArticleType}{Rapid Communication\xspace}

\newcommand{\HFK}{\mathpzc{H}_{\,\mathrm{FK}}}
\newcommand{\Htot}{\hat{{H}}}
\newcommand{\Hg}{\hat{{H}}_{g}}
\newcommand{\Ht}{\hat{{H}}_{t}}
\newcommand{\Hsigma}{\hat{H}_{g,\sigma}}
\newcommand{\Htau}{\hat{H}_{g,\tau}}
\newcommand{\Hmu}{\hat{H}_{g,\mu}}

\newcommand{\up}{\uparrow}
\newcommand{\dw}{\downarrow}
\newcommand{\etal}{\textit{et al}}
\newcommand{\ncm}[1]{n^{}_{\mathrm{CM}}(#1)}
\newcommand{\nrm}[1]{n^{}_{\mathrm{RM}}(#1)}
\newcommand{\ndual}[1]{\tilde{n}^{}_{\mathrm{CM}}(#1)}
\newcommand{\muF}{\mu^{}_\textrm{F}}

\begin{document}
\let\emph\textit

\title{%
  Majorana stripe order on the surface of a three-dimensional topological insulator
}%

\author{Y. Kamiya}
\email{yoshitomo.kamiya@riken.jp}
\affiliation{Condensed Matter Theory Laboratory, RIKEN, Wako, Saitama 351-0198, Japan}
\author{A. Furusaki}
\affiliation{Condensed Matter Theory Laboratory, RIKEN, Wako, Saitama 351-0198, Japan}
\author{J.~C.~Y. Teo}
\affiliation{Department of Physics, University of Virginia, Charlottesville, VA 22904, USA}
\author{G.-W. Chern}
\email{gchern@virginia.edu}
\affiliation{Department of Physics, University of Virginia, Charlottesville, VA 22904, USA}

\date{\today}

\begin{abstract}
  The issue on the effect of interactions in topological states concerns not only interacting topological phases but also novel symmetry-breaking phases and phase transitions. Here we study the interaction effect on Majorana zero modes (MZMs) bound to a square vortex lattice in two-dimensional (2D) topological superconductors. Under the neutrality condition, where single-body hybridization between MZMs is prohibited by an emergent symmetry, a minimal square-lattice model for MZMs can be faithfully mapped to a quantum spin model, which has no sign problem in the world-line quantum Monte Carlo simulation. Guided by an insight from a further duality mapping, we demonstrate that the interaction induces a \emph{Majorana stripe state}, a gapped state spontaneously breaking lattice translational and rotational symmetries, as opposed to the previously conjectured topological quantum criticality. Away from neutrality, a mean-field theory suggests a quantum critical point induced by hybridization.
\end{abstract}

\maketitle

Topological states are currently the focus of intensive research~\cite{HasanKane10,QiZhangreview11,Chiu2016}. In particular, bulk-boundary correspondence is a central guiding principle, which predicts low-energy modes at the interface between topologically distinct states. It also applies to topological defects (such as dislocations and superconducting vortices) in topological matter, because they can be regarded as generalized interfaces bordering on normal states \cite{RanZhangVishwanath2009,Teo2010,Juricic2012,Slager2012,HughesYaoQi13,TeoHughes2013,Slager2014,TeoHughesReview2017}. Of particular interest are Majorana zero modes (MZMs) at vortices in 2D topological superconductors. Besides exploring the potential of MZMs for quantum computation \cite{Kitaev97,ChetanSimonSternFreedmanDasSarma,Kitaev2006,Fu2008,Beenakker11,Alicea2012,Stanescu_Majorana_review,LeijnseFlensberg12,elliott_franz_review,DasSarmaFreedmanNayak15}, the idea of designing lattices of Majorana fermions out of MZMs is fascinating in its own right, because the interaction between MZMs may lead to novel phases and critical phenomena \cite{Rahmani2015a,Rahmani2015b,Milsted2015,Chiu2015a,Chiu2015b,Zhu2016,Affleck2017,Hayata2017,Sannomiya2017}.

In this \ArticleType, we study a square lattice of interacting MZMs, which may emerge at vortices in 2D topological superfluid and superconductor~\cite{Volovik99,ReadGreen,Ivanov,Grosfeld2006,Chiu2012,Biswas2013}, as predicted for the $A$ phase of $^3$He and Sr$_2$RuO$_4$ \cite{Sarma2006,ChungBluhmKim07,Jang_SrRuO}. For definiteness, we consider a surface of a 3D strong topological insulator subject to superconducting proximity effect, as proposed by Fu and Kane~\cite{Fu2008}. The predicted surface state resembles a spinless $p_x\pm ip_y$ superconductor; see Refs.~\onlinecite{Xu2014,Xu2015} for recent experimental progress. When an Abrikosov vortex lattice is induced by a magnetic field, MZMs are expected to emerge at vortices~\cite{Teo2010}, leading to a lattice of Majorana fermions at low energies. Here we assume additional conditions to stabilize a square vortex lattice such as strong fourfold lattice anisotropy, which is less common than a triangular lattice but possible (e.g., LuNi$_2$B$_2$C \cite{Wilde1997}). We demonstrate that a faithful spin representation of a minimal interacting Hamiltonian for Majorana fermions can be derived in the square lattice under the neutrality condition, which furthermore allows for employing a quantum Monte Carlo (QMC) method \cite{GKW2016} to investigate thermodynamic properties of very large lattices unbiasedly. We find a novel \emph{Majorana stripe phase} and present a duality transformation elucidating the nature of this phase, which supersedes the previously proposed topological quantum criticality~\cite{Chiu2015a}. We then extend our analysis away from neutrality by a mean-field (MF) theory by including the nearest-neighbor hybridization, where we find a quantum critical point induced by Majorana hybridization, beyond which Majorana fermions have gapless dispersion.

At the non-interacting level, the system is described in the long-wavelength limit by the Fu-Kane Hamiltonian \cite{Fu2008}
$\hat{H}_{\,\textrm{FK}} = \frac{1}{2} \int d^2\mathbf{r}\, \hat{\Psi}^\dag_\mathbf{r} \HFK(\mathbf{r}) \hat{\Psi}^{\;}_\mathbf{r}$ 
with
$\hat{\Psi}^{\;}_\mathbf{r} = (\hat{\psi}^{\;}_{\up \mathbf{r}},\hat{\psi}^{\;}_{\dw \mathbf{r}},\hat{\psi}^{\dag}_{\dw \mathbf{r}},-\hat{\psi}^{\dag}_{\up \mathbf{r}})^\mathrm{T}$
being the Nambu spinor of the electronic operators $\hat{\psi}^{(\dag)}_{\alpha\mathbf{r}}$ ($\alpha = \up,\dw$) and
\begin{align}
  \!\!\!\!
  \HFK(\mathbf{r})
  = {\tau}^z\! \left(-i v^{}_\textrm{F}{\bm{\sigma}}\cdot\nabla - \muF\right)
  + \mathrm{Re}\,\Delta(\mathbf{r}) {\tau}^x
  + \mathrm{Im}\,\Delta(\mathbf{r}) {\tau}^y,
  \label{eq:HFK}
\end{align}
where ${\sigma}$ (${\tau}$) is the Pauli matrix in the spin (Nambu) basis, $\muF$ is the chemical potential, $\Delta({\bf r})$ is the proximity-induced pair potential, and $v^{}_\textrm{F}$ is velocity of the surface Dirac mode when $\Delta = 0$. The distribution and structure of vortices are encoded in $\Delta(\mathbf{r})$. The neutrality condition corresponds to $\muF = 0$, which has a significant consequence on the emergent symmetry of the effective Hamiltonian~\cite{Chiu2015a,Chiu2015b}. When satisfied, an artificial time-reversal symmetry $\Theta_\textrm{eff} = {\sigma}^x {\tau}^x K$ ($K$ is the complex conjugation) with $\Theta_\textrm{eff}^2 = 1$ emerges in addition to the particle-hole symmetry $\Xi = {\sigma}^y {\tau}^y K$ inherent to the Bogoliubov-de Gennes formalism. The consequence is that the vortex-bound MZM takes the form,
$\hat{\gamma}^{} = \hat{\gamma}^{\dag} = \int d^2 \mathbf{r}\bigl( u^{}_{\mathbf{r}} \hat{\psi}^{\;}_{\dw,\mathbf{r}} + u^{\ast}_{\mathbf{r}} \hat{\psi}^{\dag}_{\dw,\mathbf{r}} \bigr)$, i.e., with the spin antiparallel to the magnetic field \cite{Chiu2015b}. Because
$\hat{\Theta}_\textrm{eff} \,\hat{\gamma}\, \hat{\Theta}^{-1}_\textrm{eff} = \hat{\gamma}$
and $\hat\Theta_{\mathrm{eff}}$ is antiunitary, single-body hybridization $i \hat{\gamma}^{}_{\mathbf{r}} \hat{\gamma}^{}_{\mathbf{r}'}$ is prohibited between any pair of MZMs at $\mathbf{r}$ and $\mathbf{r}'$. For an interacting many-body system, this means that the neutrality condition corresponds to the strong-coupling limit for the Majorana modes.

\begin{figure}
  \centering
  \includegraphics[width=0.95\hsize,bb=0 0 627 482]{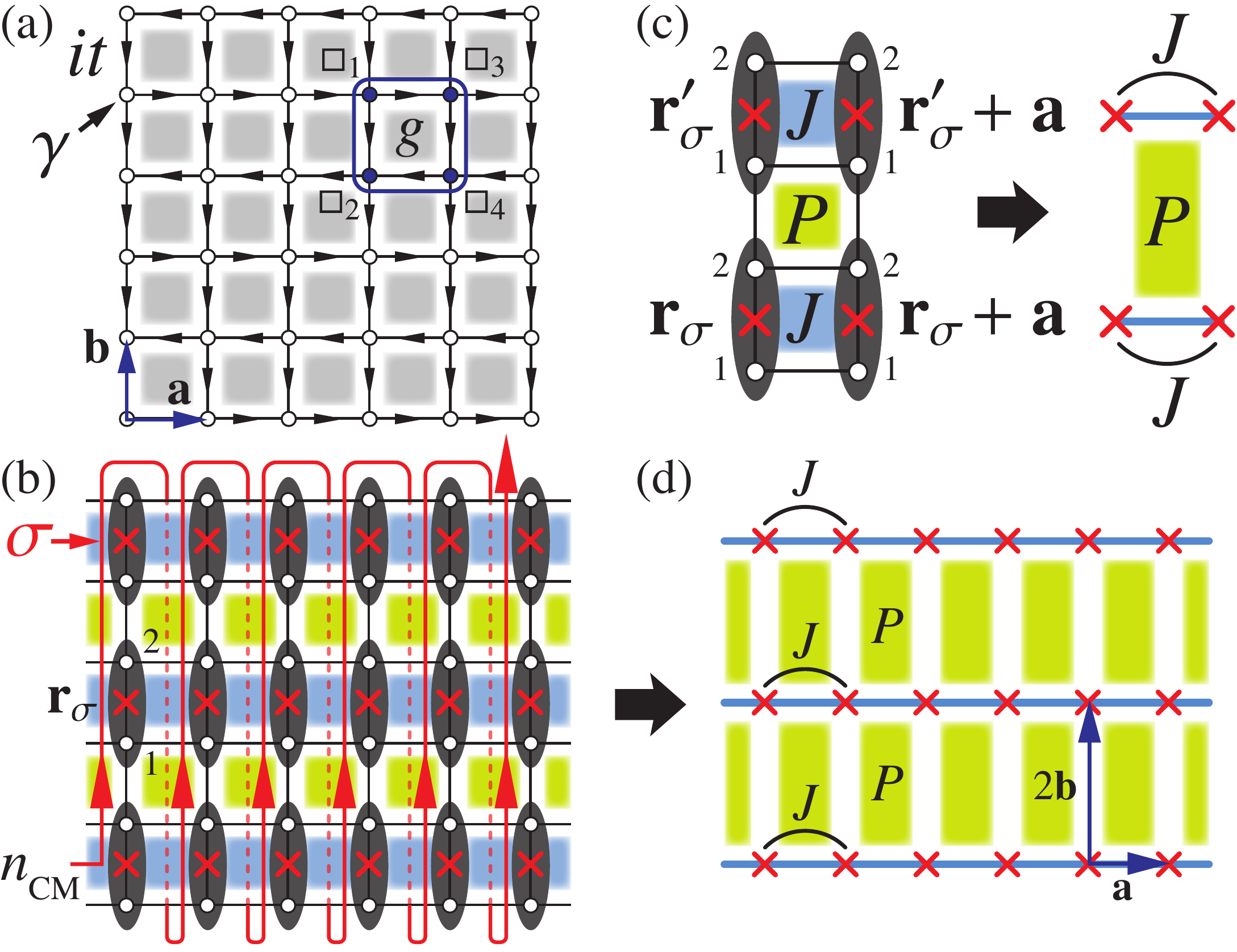}
  \caption{%
    \label{fig:model}
    MZMs and the JW transformation.
    (a) Square lattice of MZM, where shaded plaquettes represent the interaction $g$ and directed links from site $\mathbf{r}$ to site $\mathbf{r}'$ represent the hybridization term $i t \hat{\gamma}^{}_{\mathbf{r}} \hat{\gamma}^{}_{\mathbf{r}'}$.
    (b) JW transformation of $\Hg$ with the column-major site ordering $n^{}_\textrm{CM}$. Ellipses show pairing of MZMs with crosses representing $\sigma$ spins.
    (c) A plaquette term involving two (four) pairs is transformed to the Ising (four-spin) coupling, where $\mathbf{r}_\sigma' = \mathbf{r}_\sigma + 2\mathbf{b}$.
    (d) Lattice of $\sigma$ spins.
  }
\end{figure}

Assuming the simplest, quartic local interaction of the vortex Majorana modes on the square lattice, we consider the following Hamiltonian $\Htot = \Hg$ \cite{Chiu2015a,Affleck2017} with
\begin{align}
  \Hg = g \sum_{\square} \hat{\gamma}^{}_{\square_1} \hat{\gamma}^{}_{\square_2} \hat{\gamma}^{}_{\square_3} \hat{\gamma}^{}_{\square_4},
  \label{eq:Hg}
\end{align}
where
$\hat{\gamma}^{}_{\mathbf{r}}$ is the Majorana fermion operator at site $\mathbf{r}$ satisfying
$\hat{\gamma}^{\dag}_{\mathbf{r}} = \hat{\gamma}^{}_{\mathbf{r}}$ and $\{\hat{\gamma}^{}_{\mathbf{r}},\hat{\gamma}^{}_{\mathbf{r}'}\} = 2 \delta_{\mathbf{r},\mathbf{r}'}$
and the summation runs over elementary plaquettes; $\square_1$--$\square_4$ are four corners of a plaquette, $\square_2 = \square_1 - \mathbf{b}$, $\square_3 = \square_1 + \mathbf{a}$, and $\square_4 = \square_3 - \mathbf{b}$, with $\mathbf{a}$ and $\mathbf{b}$ the primitive lattice vectors [Fig.~\ref{fig:model}(a)].

As the hybridization term allowed for $\muF \ne 0$~\cite{Franz2000}, we consider the following nearest-neighbor hybridization, discussed in, e.g.,  Refs.~\onlinecite{Chiu2015a,Grosfeld2006},
\begin{align}
  \Ht
  = i t \sum_{\mathbf{r}}
  \left[
    \hat{\gamma}^{\;}_{\mathbf{r}} \hat{\gamma}^{\;}_{\mathbf{r} - \mathbf{b}}
  + (-1)^{r_y} \hat{\gamma}^{}_{\mathbf{r}} \hat{\gamma}^{}_{\mathbf{r} + \mathbf{a}}
  \right],
  \label{eq:Ht}
\end{align}
which has a uniform $\pi$ flux per plaquette because of the underlying vortices. Equation~\eqref{eq:Ht} preserves the full symmetry of the square lattice (e.g., the translation in the $\mathbf{b}$ direction is accompanied by a gauge transformation). By continuity, we expect $\lvert{t}\rvert \ll \lvert{g}\rvert$ for small $\muF$. We assume $g,t > 0$ unless otherwise mentioned.

We start from $\Htot = \Hg$~\eqref{eq:Hg}. Assuming a periodic (open) boundary condition in the $\mathbf{a}$ ($\mathbf{b}$) direction, we map it to a spin model by using a 2D Jordan-Wigner (JW) transformation. We define a complex fermion
$\hat{c}_{\mathbf{r}_\sigma}^{} = \frac{1}{2}(\hat{\gamma}^{}_{\mathbf{r}_\sigma,1} + i \hat{\gamma}^{}_{\mathbf{r}_\sigma,2})$
by introducing an artificial (but arbitrary) pairing convention [Fig.~\ref{fig:model}(b)], where $\mathbf{r}_\sigma$ is the position of a pair combining $\hat{\gamma}^{}_{\mathbf{r}_\sigma,1}$ and $\hat{\gamma}^{}_{\mathbf{r}_\sigma,2}$. Assuming the site-ordering (``column-major'') index $\ncm{\mathbf{r}_\sigma}$ in Fig.~\ref{fig:model}(b), the transformation is
$\hat{c}^{\dag}_{\mathbf{r}_\sigma} \hat{c}^{\;}_{\mathbf{r}_\sigma} = \frac{1}{2} \bigl( 1 + \hat{\sigma}_{\mathbf{r}_\sigma}^z \bigr)$
and
$
\hat{c}^{\dag}_{\mathbf{r}_\sigma}
= \frac{1}{2} \bigl( \prod_{\ncm{\mathbf{r}_\sigma'} < \ncm{\mathbf{r}_\sigma}} \hat{\sigma}_{\mathbf{r}_\sigma'}^z \bigr)
\bigl( \hat{\sigma}_{\mathbf{r}_\sigma}^x + i \hat{\sigma}_{\mathbf{r}_\sigma}^y \bigr)
$,
where $\hat{\sigma}^\alpha$ ($\alpha = x,y,z$) are Pauli matrices. We find
\begin{align}
  \hat{\gamma}^{}_{\mathbf{r}_\sigma,2}
  \hat{\gamma}^{}_{\mathbf{r}_\sigma,1}
  \hat{\gamma}^{}_{\mathbf{r}_\sigma + \mathbf{a},2}
  \hat{\gamma}^{}_{\mathbf{r}_\sigma + \mathbf{a},1}
  &= - \hat{\sigma}_{\mathbf{r}_1}^z \hat{\sigma}_{\mathbf{r}_1 + \mathbf{a}}^z,
  \notag\\
  \hat{\gamma}^{}_{\mathbf{r}_\sigma',1}
  \hat{\gamma}^{}_{\mathbf{r}_\sigma,2}
  \hat{\gamma}^{}_{\mathbf{r}_\sigma' + \mathbf{a},1}
  \hat{\gamma}^{}_{\mathbf{r}_\sigma + \mathbf{a},2}
  &= - \hat{\sigma}_{\mathbf{r}_\sigma}^x \hat{\sigma}_{\mathbf{r}_\sigma'}^x \hat{\sigma}_{\mathbf{r}_\sigma + \mathbf{a}}^x \hat{\sigma}_{\mathbf{r}_\sigma' + \mathbf{a}}^x,
\end{align}
with $\mathbf{r}_\sigma' = \mathbf{r}_\sigma + 2\mathbf{b}$, where the number of pairs involved in the interaction is two and four, respectively [Fig.~\ref{fig:model}(c)]. The string factor does not appear in either case. We obtain
\begin{align}
  \Hsigma
  = -J \sum_{\mathbf{r}_\sigma} \hat{\sigma}_{\mathbf{r}_\sigma}^z \hat{\sigma}_{\mathbf{r}_\sigma+\mathbf{a}}^z
  - P \sum_{\square_\sigma} \left( \prod_{\mathbf{r}_\sigma \in \square_\sigma} \hat{\sigma}_{\mathbf{r}_\sigma}^x \right),  
  \label{eq:Hspin}
\end{align}
with $J = P = g$, which combines the Ising coupling $J$ on the horizontal bonds and a transverse four-spin term $P$ associated with plaquettes ($\square_\sigma$) of $\sigma$ spins [Fig.~\ref{fig:model}(b)]. In this representation~\eqref{eq:Hspin}, we can apply the world-line QMC method \cite{GKW2016} to study the thermodynamic properties of MZMs without a negative sign problem. Specifically, we use the directed-loop algorithm \cite{Syljuasen2001,Alet2005} in the $\sigma^x$ basis. To reduce finite-size effects, we use a trick of fictitious MZMs to simulate the lattice of Majorana fermions comprising an even number of plaquettes in the $\mathbf{b}$ direction \cite{SM}. We investigate the spin lattices of $L \times L$ up to $L = 60$, corresponding to $L \times (2 L - 1)$ MZMs, significantly larger than the previous exact diagonalization (ED) study up to $4 \times 15$ MZMs~\cite{Chiu2015a}.

\begin{figure*}
  \centering
  \includegraphics[width=1.00\hsize,bb=0 0 1657 448]{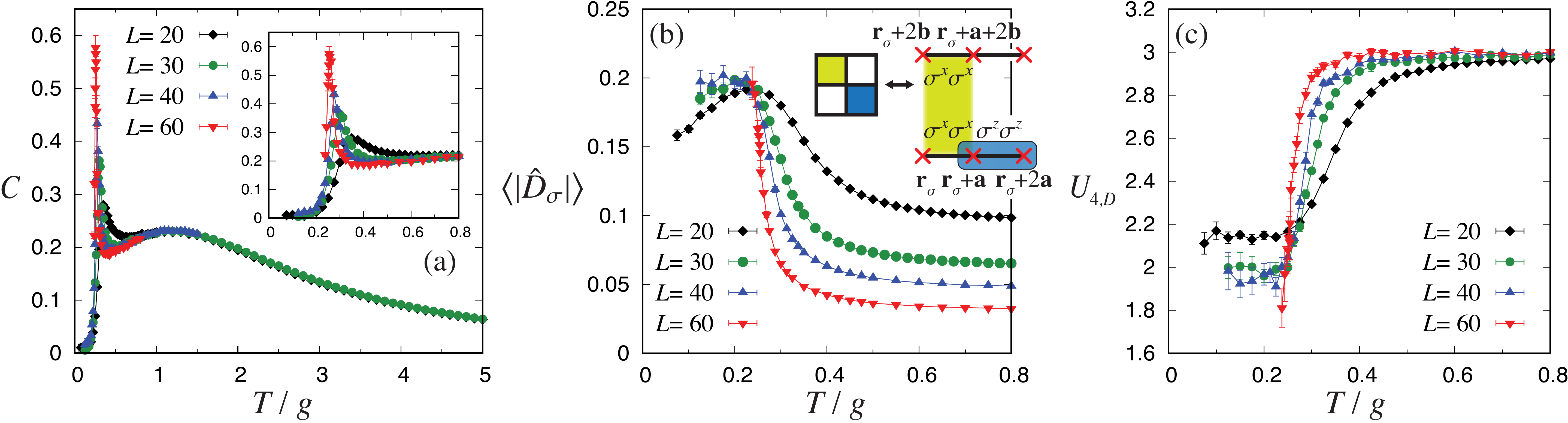}
  \caption{%
    \label{fig:qmc}
    QMC results in the spin representation~\eqref{eq:Hspin} with $J = P = g$ in the $L \times L$ lattice with the fictitious MZM trick \cite{SM}.
    (a) Specific heat
    $C = (1/L^{2})\partial\langle{\Hsigma}\rangle/\partial T$,
    (b) the order parameter $\langle{\lvert{\hat{D}_\sigma}\rvert}\rangle$ of the Majorana stripe state,
    and
    (c) the Binder parameter. The inset in (b) illustrates the local order parameter $\hat{D}_\sigma({\mathbf{r}_\sigma})$~\eqref{eq:D} and its relation with a pair of Majorana plaquettes.
  }
\end{figure*}

Figure~\ref{fig:qmc}(a) shows the specific heat $C$. In addition to the broad peak around temperature $T \approx g$, it exhibits a size-dependent sharp anomaly at $T / g \approx 0.25$, indicating a finite-temperature transition. This observation points to a symmetry breaking phase at low $T$, which contradicts with the previous ED study, where by introducing a two-site modulation in $g$ [equivalent to making $J \ne P$ in Eq.~\eqref{eq:Hspin}], it was suggested that the system becomes gapless when the original translational and rotational symmetries are recovered (i.e., $J = P$)~\cite{Chiu2015a}. To clarify the nature of the low-$T$ state, we first note that the string operator $\hat{O}^\textrm{spin}_\textrm{h}(y) = \prod_{r^x_\sigma} \hat{\sigma}^x_{\mathbf{r}_\sigma = (r^x_\sigma,y)}$ is a conserved quantity for any horizontal ($\parallel \mathbf{a}$) chain, which flips $\sigma^z$ eigenvalues of all spins at $r^y_\sigma = y$. As known as a generalized Elitzur's theorem \cite{Batista2005}, the corresponding gauge-like 1D symmetries reduce the effective dimensionality of the order parameter field $\sigma^z$ from 2D to 1D. Hence, the conservation of $\hat{O}^\textrm{spin}_\textrm{h}(y)$ prohibits any kind of long-range order of $\sigma^z$ at $T > 0$; this 1D physics may explain the broad peak of $C$ at high $T$, but not the transition itself.

\begin{figure}
  \centering
  \includegraphics[width=\hsize,bb=0 0 833 677]{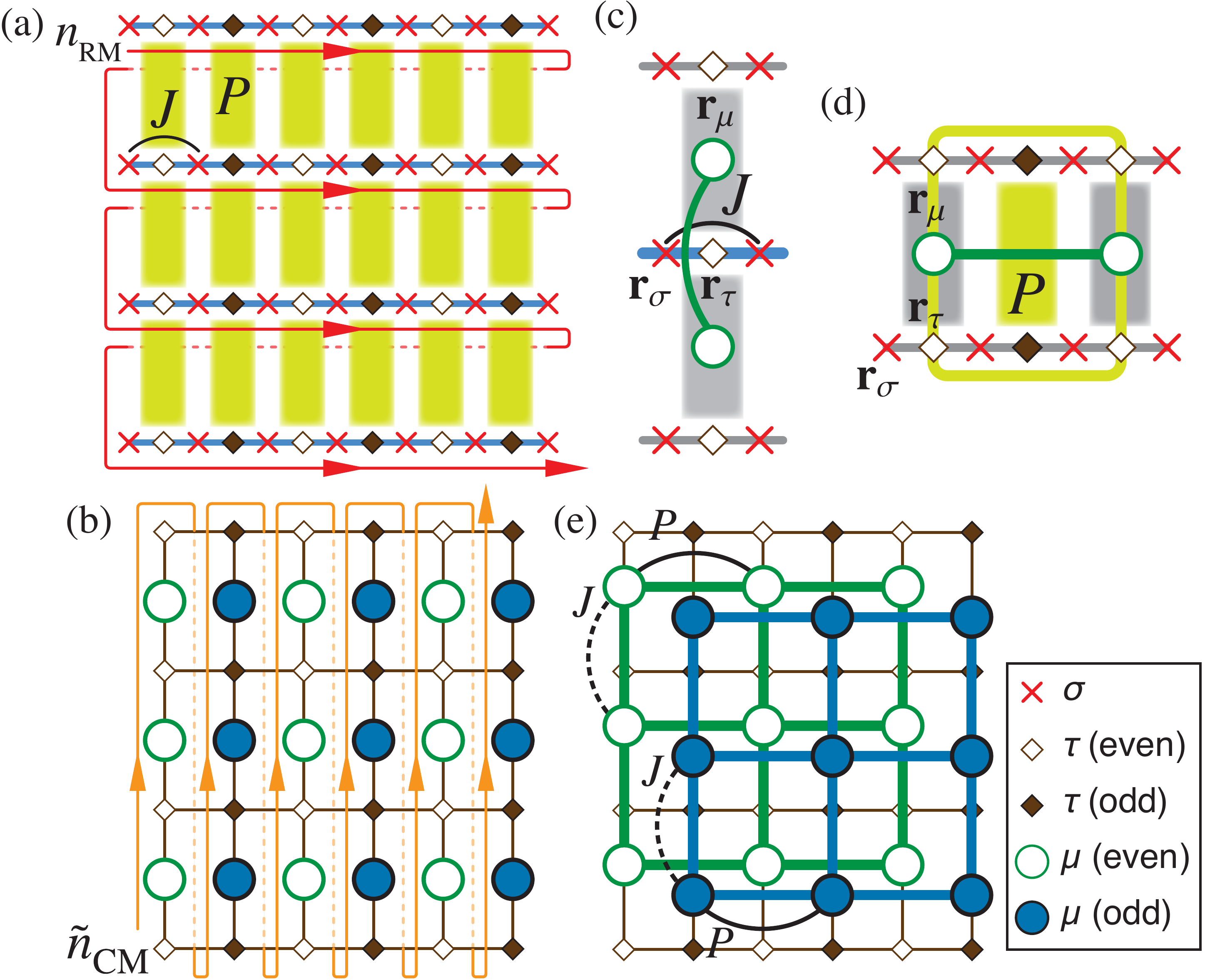}
  \caption{%
    \label{fig:duality}
    Two-step duality transformation, introducing
    (a) $\tau$ spins and (b) $\mu$ spins.
    (c) The $J$ term is transformed to the interaction
    $\hat{\mu}^z_{\mathbf{r}_\mu}\hat{\mu}^z_{\mathbf{r}_\mu + 2\mathbf{b}}$.
    (d) The $P$ term (highlighted filled rectangle) is transformed first into a four-spin coupling for $\tau$ spins (round rectangle) and then to
    $\hat{\mu}^x_{\mathbf{r}_\mu} \hat{\mu}^x_{\mathbf{r}_\mu + 2\mathbf{a}}$.
    (e) Resulting decoupled copies of the quantum compass model for $\mu$ spins (shifted vertically for clarity).
  }
\end{figure}

To elucidate the nature of the low-$T$ phase and the transition, we show that $\Hsigma$ (hence, $\Hg$) is dual to decoupled two copies of a square-lattice quantum compass model \cite{SM}. This model was investigated in depth in various contexts \cite{Nussinov2015,Mishra2004,Dorier2005,Chen2007,Tanaka2007,Wenzel2008,Orus2009,Xu2004,Xu2005,Nussinov2005,Vidal2009,Nasu2017}, and the corresponding knowledge is very useful for understanding the nature of the low-$T$ phase. Explicitly, we first define ``$\tau$ spins'' at the midpoint of every  horizontal link. With the ``row-major'' site ordering $\nrm{\mathbf{r}_\sigma}$ in Fig.~\ref{fig:duality}(a), the first transformation is
$\hat{\tau}^z_{\mathbf{r}_\tau} = \hat{\sigma}^z_{\mathbf{r}_\sigma} \hat{\sigma}^z_{\mathbf{r}_\sigma+\mathbf{a}}$,
$\hat{\tau}^x_{\mathbf{r}_\tau} = \prod_{\nrm{\mathbf{r}_\sigma'} \le \nrm{\mathbf{r}_\sigma}} \hat{\sigma}^x_{\mathbf{r}_\sigma'}$
with $\mathbf{r}_\tau = \mathbf{r}_\sigma + \frac{\mathbf{a}}{2}$, by which the $J$ and $P$ terms become an effective magnetic field and a four-spin interaction for $\tau$ spins, respectively. We find that the new four-spin interaction does not mix $\tau$ spins in even and odd columns, e.g.,
$\hat{\sigma}^x_{\mathbf{r}_\sigma} \hat{\sigma}^x_{\mathbf{r}_\sigma+\mathbf{a}}\hat{\sigma}^x_{\mathbf{r}_\sigma + 2\mathbf{b}} \hat{\sigma}^x_{\mathbf{r}_\sigma+\mathbf{a}+2\mathbf{b}} = \hat{\tau}^x_{\mathbf{r}_\tau-\mathbf{a}} \hat{\tau}^x_{\mathbf{r}_\tau+\mathbf{a}} \hat{\tau}^x_{\mathbf{r}_\tau-\mathbf{a}+2\mathbf{b}} \hat{\tau}^x_{\mathbf{r}_\tau+\mathbf{a}+2\mathbf{b}}$
[Fig.~\ref{fig:duality}(d)]. Consequently, the dual Hamiltonian is composed of decoupled even and odd components as $\Htau^{} = \Htau^{\mathrm{e}} + \Htau^{\mathrm{o}}$ with
\begin{align}
  \Htau^{\,\mathrm{e(o)}}
  ~ = ~
  \sum_{\mathclap{\mathbf{r}_\tau \,\in\, \text{even (odd) columns}}}
  ~
  \left(
  -J \hat{\tau}^z_{\mathbf{r}_\tau}
  -P \hat{\tau}^x_{\mathbf{r}_\tau} \hat{\tau}^x_{\mathbf{r}_\tau + 2\mathbf{a}}
  \hat{\tau}^x_{\mathbf{r}_\tau + 2\mathbf{b}} \hat{\tau}^x_{\mathbf{r}_\tau + 2\mathbf{a} + 2\mathbf{b}}
  \right).
\end{align}
To complete the mapping, we introduce ``$\mu$ spins'' at the midpoint of each vertical link $(\mathbf{r}_\tau, \mathbf{r}_\tau + 2\mathbf{b})$ for $\tau$ spins, such that
$\hat{\mu}^x_{\mathbf{r}_\mu} = \hat{\tau}^x_{\mathbf{r}_\tau} \hat{\tau}^x_{\mathbf{r}_\tau+2\mathbf{b}}$,
$\hat{\mu}^z_{\mathbf{r}_\mu} = \prod_{ \ndual{\mathbf{r}_\tau'} \le \ndual{\mathbf{r}_\tau} } \hat{\tau}^z_{\mathbf{r}_\tau'}$
with $\mathbf{r}_\mu = \mathbf{r}_\tau + \mathbf{b}$, where $\ndual{\mathbf{r}_\tau}$ is the column-major ordering for $\tau$ spins [Fig.~\ref{fig:duality}(b)]. This preserves the decoupling of $\Htau^{\,\mathrm{e}}$ and $\Htau^{\,\mathrm{o}}$, transforming each into the quantum compass model on a square lattice with an enlarged unit cell [Figs.~\ref{fig:duality}(c)--\ref{fig:duality}(e)],
\begin{align}
  \Hmu^{\,\mathrm{e(o)}}
  ~ = ~
  \sum_{\mathclap{\mathbf{r}_\mu \,\in\, \text{even (odd) column}}}
  ~
  \left(
  -P \hat{\mu}^x_{\mathbf{r}_\mu} \hat{\mu}^x_{\mathbf{r}_\mu + 2\mathbf{a}}
  -J \hat{\mu}^z_{\mathbf{r}_\mu} \hat{\mu}^z_{\mathbf{r}_\mu + 2\mathbf{b}}
  \right).
  \label{eq:QCM}
\end{align}
The total Hamiltonian is $\Hmu^{} = \Hmu^{\mathrm{e}} + \Hmu^{\mathrm{o}}$.

\begin{figure}
  \centering
  \includegraphics[width=0.9\hsize, bb=0 0 804 914]{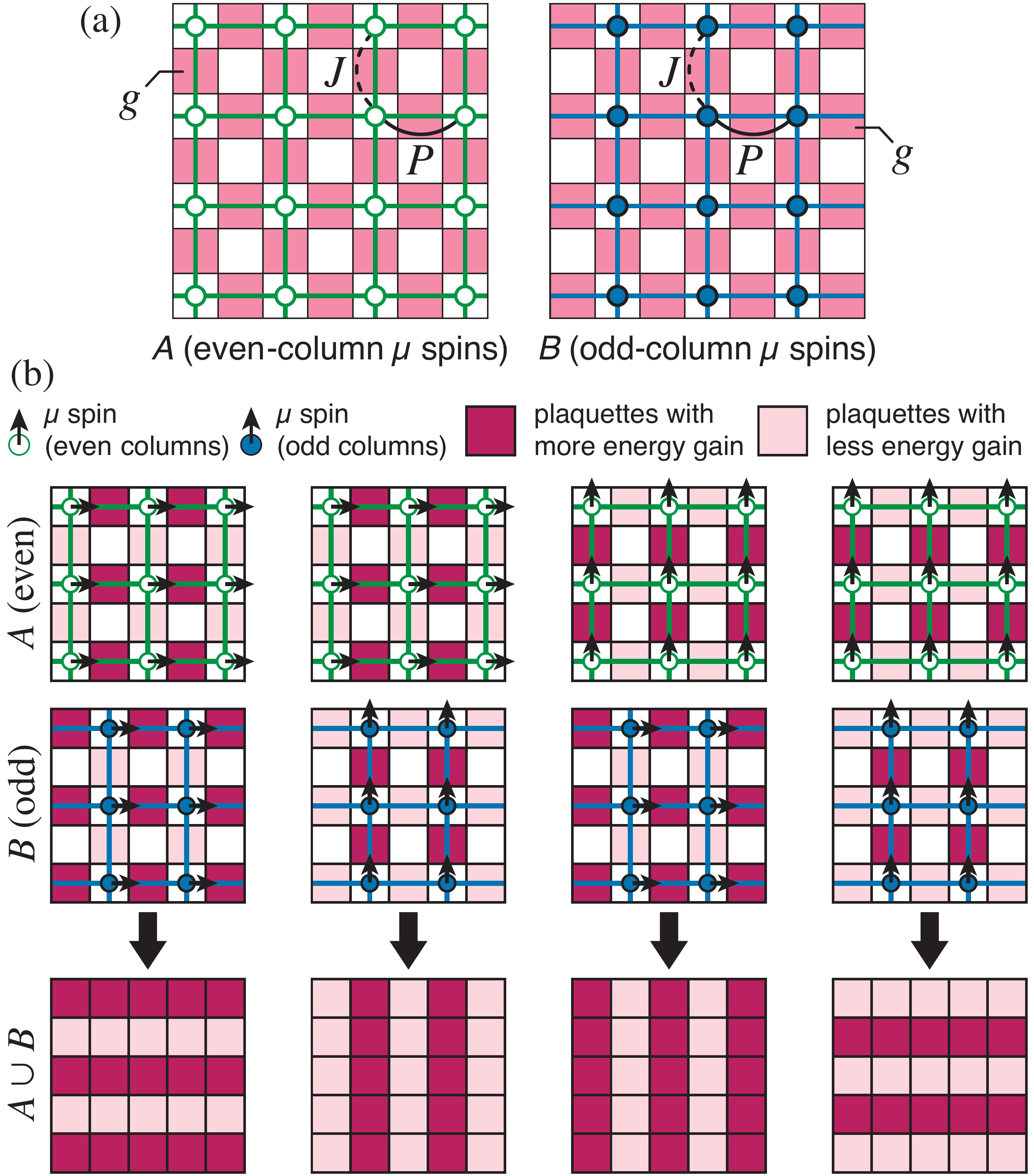}
  \caption{%
    \label{fig:stripe}
    Majorana stripe order.
    (a) Correspondence between the checkerboard decomposition of $\Hg$~\eqref{eq:Hg} and the even and odd components of the compass model.
    (b) Fourfold degenerate ordered state and the relation with the nematic order in the compass model.
  }
\end{figure}

The most crucial input from the duality transformation is that the compass model with $J = P = g$ is known to undergo a ``nematic'' transition in the Ising universality class at a finite temperature \cite{Mishra2004,Tanaka2007,Wenzel2008}. Below the critical temperature $T = T_c$, the $\mathbb{Z}_2$ spin-lattice reflection symmetry [$x \leftrightarrow z$ ($\mathbf{a} \leftrightarrow \mathbf{b}$) in the spin (real) space] is spontaneously broken, while any spin-spin correlation function such as $\langle{\hat{\mu}^x_{\mathbf{r}_\mu} \hat{\mu}^x_{\mathbf{r}_\mu'}}\rangle$ and $\langle{\hat{\mu}^z_{\mathbf{r}_\mu} \hat{\mu}^z_{\mathbf{r}_\mu'}}\rangle$ remains short-ranged. This $\mathbb{Z}_2$ symmetry breaking can be detected by a directional order parameter
$\hat{D}_{\mu}(\mathbf{r}_\mu) = \hat{\mu}^x_{\mathbf{r}_\mu} \hat{\mu}^x_{\mathbf{r}_\mu + 2\mathbf{a}} - \hat{\mu}^z_{\mathbf{r}_\mu} \hat{\mu}^z_{\mathbf{r}_\mu + 2\mathbf{b}}$~\cite{Nussinov2015}.
Back to the language of Majorana fermions, the even-odd decomposition ($\Hmu^{} = \Hmu^{\mathrm{e}} + \Hmu^{\mathrm{o}}$) corresponds to the geometrical checkerboard decomposition of $\Hg$~\eqref{eq:Hg}. Defining $\Hg^A$ and $\Hg^B$ as composed of quartic interactions in one sublattice ($A$) of the checkerboard decomposition and its complement ($B$), respectively [Fig.~\ref{fig:stripe}(a)], we find $\Hg = \Hg^A + \Hg^B$ and $[\Hg^A, \Hg^B] = 0$. Here, $\Hg^A$ corresponds to $\Hmu^{\mathrm{e}}$ or $\Hmu^{\mathrm{o}}$ and $\Hg^B$ does to the other. In fact, each Ising-like bond interaction in $\Hmu^{\,\mathrm{e(o)}}$~\eqref{eq:QCM} corresponds to a plaquette term that it graphically overlaps in the lattice, as illustrated in Fig.~\ref{fig:stripe}(a). Hence, the nematic order quantified by $\hat{D}_{\mu}$ corresponds to a \emph{spontaneous energy density modulation} associated with the plaquette interaction $g$. As shown in Fig.~\ref{fig:stripe}(b), the even-odd decoupling implies that the energy-density wave order emerges in the two sublattices $A$ and $B$ independently ($\mathbb{Z}_2 \times \mathbb{Z}_2$ symmetry breaking), resulting in fourfold degenerate ground states modulo the aforementioned 1D symmetries.

We confirm this Majorana stripe order by evaluating the order parameter by QMC. Figure~\ref{fig:qmc}(b) shows $\langle{\lvert{\hat{D}_\sigma}\rvert}\rangle$ with $\hat{D}_\sigma = \mathcal{N}^{-1} \sum'_{\mathbf{r}_\sigma} \hat{D}_\sigma({\mathbf{r}_\sigma})$, where the summation runs over either even or odd columns, $\mathcal{N}$ is a proper normalization \cite{SM}, and
\begin{align}
  \hat{D}_\sigma({\mathbf{r}_\sigma})
  =
  \hat{\sigma}_{\mathbf{r}_\sigma + \mathbf{a}}^z
  \hat{\sigma}_{\mathbf{r}_\sigma + 2\mathbf{a}}^z
  -
  \hat{\sigma}_{\mathbf{r}_\sigma + 2\mathbf{b}}^x
  \hat{\sigma}_{\mathbf{r}_\sigma}^x
  \hat{\sigma}_{\mathbf{r}_\sigma + \mathbf{a} + 2\mathbf{b}}^x
  \hat{\sigma}_{\mathbf{r}_\sigma + \mathbf{a}}^x
  \label{eq:D}
\end{align}
is the order parameter in the $\sigma$-spin representation. We find that the temperature dependence of $\langle{\lvert{D}\rvert}\rangle$ is consistent with the transition into the Majorana stripe phase (the nonmonotonic $T$-dependence for small $L$ is suggested to be a finite-size effect due to the open boundary condition in the $\mathbf{b}$ direction). Figure~\ref{fig:qmc}(c) shows the Binder parameter $U_{4,D} = \langle{\hat{D}_\sigma^4}\rangle / \langle{\hat{D}_\sigma^2}\rangle^2$, which exhibits crossing for different $L$, providing another confirmation of the transition. The crossing takes place at $T_c/g \approx 0.25(1)$, in agreement with the location of the divergent peak of $C$.

\begin{figure}
  \centering
  \includegraphics[width=0.95\hsize,bb=0 0 885 735]{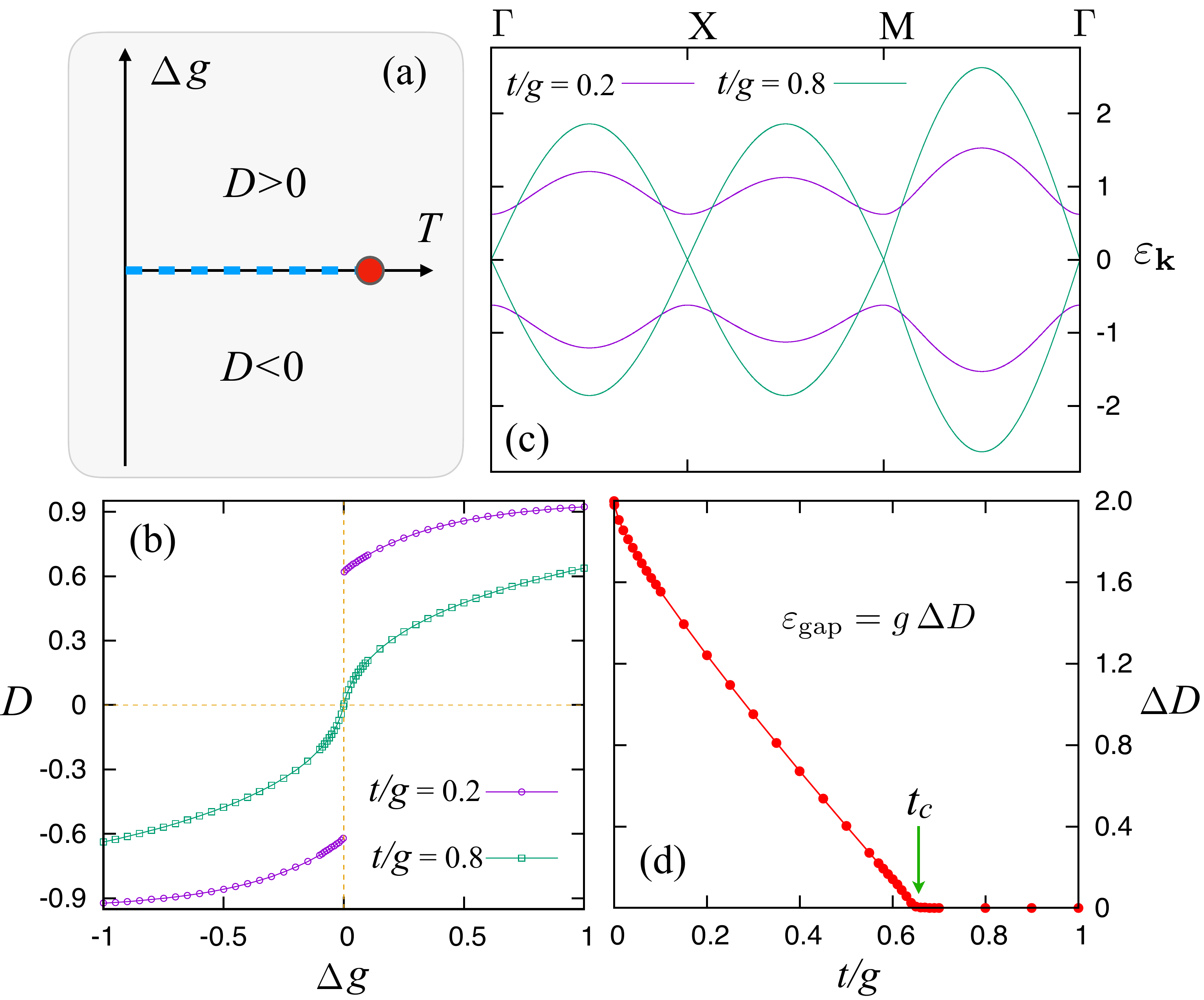}
  \caption{%
    \label{fig:MF} MF results for $\Ht + \Hg$.
    (a) Schematic phase diagram in the $T$-$\Delta g$ plane showing a first-order transition line ending at finite-$T$ critical point.
    (b) Stripe order parameter $D$ as a function of $\Delta g$ for
    $t/g=0.2, 0.8$.
    (c) Majorana MF band structure for varying $t$.
    (d) Energy gap $\varepsilon_{\rm gap}(t)$ for $\Delta g = 0$, which is related to the jump $\Delta D$ of the stripe order parameter at $\Delta g = 0$ as $\varepsilon_{\rm gap} = g \Delta D$ \cite{SM}.
  }
\end{figure}

Finally, we consider the effect of the nearest-neighbor hybridization $\Ht$~\eqref{eq:Ht} on the Majorana stripe phase. The finite-temperature Ising transition implies a first-order transition line in the extended $T$-$\Delta g$ phase diagram [Fig.~\ref{fig:MF}(a)], where nonzero $\Delta g \equiv P - J$ explicitly breaks the translational symmetry [Fig.~\ref{fig:model}(c)]. Since the QMC method has the sign problem when applied to $\Htot = \Hg + \Ht$, we employ the MF approximation in the Majorana representation to examine the discontinuous transition at $T = 0$. Figure~\ref{fig:MF}(b) shows the order parameter $D = \langle{\hat{D}_\sigma}\rangle$ as a function of $\Delta g / g$ for $t/g = 0.2, 0.8$. For $t/g = 0.2$, $D$ remains finite as $\Delta g \to 0$, and exhibits a jump upon changing the sign of $\Delta g$, indicating that the discontinuous transition persists even in the presence of weak hybridization. This in turn implies that the finite-$T$ transition remains stable for small $t$, although the induced coupling between the $A$ and $B$ subsystems may alter the universality class \cite{Ashkin1943}. As $t$ increases, the discontinuity at $\Delta g = 0$, $\Delta D$, is reduced and vanishes for $t > t_c \approx 0.65g$ [Fig.~\ref{fig:MF}(d)]. As shown in Fig.~\ref{fig:MF}(c), the MF band structure of Majorana fermions in the limit $\Delta g \to 0$ in the stripe phase ($t < t_c$) is gapped with the energy gap $\varepsilon_{\rm gap} = g \Delta D$~\cite{SM}. This band structure is topologically trivial~\cite{SM}. With increasing $t$, the gap reduces and vanishes for $t \ge t_c$, giving rise to a critical state with gapless Majorana fermions. Assuming that the critical temperature $T_c \approx \varepsilon_{\rm gap}$, our result suggests $T_c \to 0$ as $t \to t_c$. Our calculation thus points to the existence of a quantum critical point characterized by gapless Majorana fermion excitations for $t \ge t_c$. We note that the effect of including second-neighbor hybridization, which produces a gap in the excitation spectrum, was recently discussed~\cite{Affleck2017}.

In summary, the square-lattice Majorana Hamiltonian $\Ht+\Hg$, which may have an experimental realization in the hybrid of a 3D strong topological insulator and a superconductor, induces a stripe order that spontaneously breaks the translational and rotational symmetries in the strong-coupling regime $g \gg t$, as opposed to the previously conjectured topological quantum critical behavior~\cite{Chiu2015a}. Our large-scale QMC simulation as well as the duality mapping (via the JW transformation) provide a solid confirmation of this phenomenon. We note that Affleck \etal.~also investigated the same model recently from the weak-coupling side, suggesting that the quantum phase transition $t = t_c$ belongs to a supersymmetric universality class \cite{Affleck2017}. Our unbiased approach coming from the strong coupling is complementary to their weak-coupling analysis. In fact, $t \ne 0$ lifts the 1D gauge-like symmetries, reducing the Majorana stripe state to the dimerized state found by Affleck \etal.~using a MF treatment similar to ours. We hope that our work will trigger an experimental effort in the search for intriguing phase transitions in the system of interacting Majorana modes, which may be synthesized on the surface of a 3D topological insulator as proposed recently~\cite{Chiu2015a,Chiu2015b}.

\begin{acknowledgments}
  We are grateful to Sharmistha Sahoo for useful discussions at the early stage of this project and to Cristian Batista for valuable discussions. The numerical simulations in this work in part utilized the facilities of the Supercomputer Center, ISSP, the University of Tokyo. Y.K.~acknowledges support by JSPS Grants-in-Aid for Scientific Research under Grant No.~JP16H02206. A.F.~acknowledges support by JSPS Grants-in-Aid for Scientific Research under Grant No.~15K05141. J.C.Y.T.~is supported by the NSF under Grant No.~DMR-1653535.
\end{acknowledgments}

%\bibliographystyle{apsrev4-1.bst}
%\bibliography{ref}

%merlin.mbs apsrev4-1.bst 2010-07-25 4.21a (PWD, AO, DPC) hacked
%Control: key (0)
%Control: author (72) initials jnrlst
%Control: editor formatted (1) identically to author
%Control: production of article title (-1) disabled
%Control: page (0) single
%Control: year (1) truncated
%Control: production of eprint (0) enabled
%

%% Inclusion of Supplemental Material for an arXiv preprint
\if\arXiv1
\vspace{10pt}
\begin{center}
  ------
\end{center}
\vspace{10pt}

\appendix
\begin{center}
{\large\bf Supplemental Material}
\end{center}

%\documentclass[aps,prl,10pt,twocolumn,footnoteinbib,superscriptaddress]{revtex4-1}
%\begin{document}
%\newcommand{\temp}{\vspace{10pt}\noindent------\\------\\\vspace{10pt}}

% Supplemental Materials
\setcounter{figure}{0}
\setcounter{equation}{0}
\setcounter{table}{0}
\renewcommand{\thefigure}{S\arabic{figure}}
\renewcommand{\theequation}{S\arabic{equation}}
\renewcommand{\thetable}{S\Roman{table}}

\section{Two-step duality transformation}
Here we provide some details of the two-step duality transformation that we invoked to derive the compass model from the effective spin model $\hat{{H}}_{g,\sigma}$ [Eq.~(5) in the main text] for the square-lattice Majorana fermions under the neutrality condition.
$\hat{{H}}_{g,\sigma}$ is defined in terms of ``$\sigma$ spins'' obtained by the Jordan-Wigner transformation. They are represented by Pauli matrices (eigenvalue $\pm 1$) $\hat{\sigma}_{\mathbf{r}_\sigma}^{x}$, $\hat{\sigma}_{\mathbf{r}_\sigma}^{y}$, and $\hat{\sigma}_{\mathbf{r}_\sigma}^{z}$ residing at sites represented by crosses in Fig.~3(a) in the main text.

The first step is to introduce ``$\tau$ spins'' represented by Pauli matrices $\hat{\tau}_{\mathbf{r}_\tau}^{x,y,z}$, which reside at the midpoint $\mathbf{r}_\tau = \mathbf{r}_\sigma + \frac{\mathbf{a}}{2}$ of every horizontal link $(\mathbf{r}_\sigma, \mathbf{r}_\sigma+\mathbf{a})$ of a pair of $\sigma$ spins [diamonds in Fig.~3(a) in the main text]. We assume an open (a periodic) boundary condition in the horizontal (vertical) direction.
With the ``row-major'' site ordering $n^{}_{\mathrm{RM}}(\mathbf{r}_\sigma)$ for $\sigma$ spins, the first transformation is
\begin{align}
        \hat{\tau}^z_{\mathbf{r}_\tau}
        &= \hat{\sigma}^z_{\mathbf{r}_\sigma} \hat{\sigma}^z_{\mathbf{r}_\sigma+\mathbf{a}},
        \notag\\
        \hat{\tau}^x_{\mathbf{r}_\tau} &=
        \hat{Q}^{\sigma}_{\mathbf{r}_\sigma} \hat{\sigma}^x_{\mathbf{r}_\sigma},
        \notag\\
        \hat{\tau}^y_{\mathbf{r}_\tau}
        &=
        i \hat{\tau}^x_{\mathbf{r}_\tau} \hat{\tau}^z_{\mathbf{r}_\tau}
        =
        \hat{Q}^{\sigma}_{\mathbf{r}_\sigma} \hat{\sigma}^y_{\mathbf{r}_\sigma} \hat{\sigma}^z_{\mathbf{r}_\sigma+\mathbf{a}},
\end{align}
where
\begin{align}
  \hat{Q}^{\sigma}_{\mathbf{r}_\sigma}
  = \prod_{n^{}_{\mathrm{RM}}({\mathbf{r}_\sigma'}) < n^{}_{\mathrm{RM}}({\mathbf{r}_\sigma})} \hat{\sigma}^x_{\mathbf{r}_\sigma'}.
\end{align}
This is a standard Kramers-Wannier transformation.
Noting that no spin operators at site $\mathbf{r}_\sigma, \mathbf{r}_\sigma + \mathbf{a}, \dots$ are contained in $\hat{Q}_{\mathbf{r}_\sigma}$, we can immediately check the spin commutation relations for $\tau$ spins as follows:
\begin{align}
  [ \hat{\tau}^z_{\mathbf{r}_\tau}, \hat{\tau}^x_{\mathbf{r}_\tau} ]
  &= \hat{Q}^{\sigma}_{\mathbf{r}_\sigma} [ \hat{\sigma}^z_{\mathbf{r}_\sigma}, \hat{\sigma}^x_{\mathbf{r}_\sigma} ] \hat{\sigma}^z_{\mathbf{r}_\sigma+\mathbf{a}}
  =  \hat{Q}^{\sigma}_{\mathbf{r}_\sigma} \left( 2 i \hat{\sigma}^y_{\mathbf{r}_\sigma} \right) \hat{\sigma}^z_{\mathbf{r}_\sigma+\mathbf{a}}
  = 2 i \hat{\tau}^y_{\mathbf{r}_\tau},
  \notag\\
  [ \hat{\tau}^x_{\mathbf{r}_\tau}, \hat{\tau}^y_{\mathbf{r}_\tau} ]
  &= [ \hat{\sigma}^x_{\mathbf{r}_\sigma}, \hat{\sigma}^y_{\mathbf{r}_\sigma} ] \hat{\sigma}^z_{\mathbf{r}_\sigma+\mathbf{a}}
  = 2 i \hat{\sigma}^z_{\mathbf{r}_\sigma} \hat{\sigma}^z_{\mathbf{r}_\sigma+\mathbf{a}}
  = 2 i \hat{\tau}^z_{\mathbf{r}_\tau},
  \notag\\
  [ \hat{\tau}^y_{\mathbf{r}_\tau}, \hat{\tau}^z_{\mathbf{r}_\tau} ]  
  &= \hat{Q}^{\sigma}_{\mathbf{r}_\sigma} [ \hat{\sigma}^y_{\mathbf{r}_\sigma}, \hat{\sigma}^z_{\mathbf{r}_\sigma} ]
  = \hat{Q}^{\sigma}_{\mathbf{r}_\sigma} \left( 2 i \hat{\sigma}^x_{\mathbf{r}_\sigma} \right)
  = 2i \hat{\tau}^x_{\mathbf{r}_\tau}.
\end{align}
It is also straightforward to verify that $\tau$ spin operators at different sites commute.

The second step to complete the transformation is to introduce ``$\mu$ spins.'' This is another Kramers-Wannier transformation though with the ``column-major'' ordering $\tilde{n}^{}_{\mathrm{CM}}({\mathbf{r}_\sigma})$ illustrated in Fig.~3(b) in the main text. We here assume a different boundary condition, i.e., a periodic (an open) boundary condition in the vertical (horizontal) direction. The fact that we assume different boundary conditions does not matter since we only discuss thermodynamic properties. By introducing $\mu$ spins at the midpoint $\mathbf{r}_\mu = \mathbf{r}_\tau + \mathbf{b}$ of each vertical link $(\mathbf{r}_\tau, \mathbf{r}_\tau + 2\mathbf{b})$ for $\tau$ spins, the transformation is
\begin{align}
  \hat{\mu}^x_{\mathbf{r}_\mu} &= \hat{\tau}^x_{\mathbf{r}_\tau} \hat{\tau}^x_{\mathbf{r}_\tau+2\mathbf{b}},
  \notag\\
  \hat{\mu}^z_{\mathbf{r}_\mu} &=
  \hat{Q}^{\tau}_{\mathbf{r}_\tau} \hat{\tau}^z_{\mathbf{r}_\tau},
  \notag\\
  \hat{\mu}^y_{\mathbf{r}_\mu} &= i \hat{\mu}^x_{\mathbf{r}_\mu} \hat{\mu}^z_{\mathbf{r}_\mu}
  = \hat{Q}^{\tau}_{\mathbf{r}_\tau}
  \hat{\tau}^y_{\mathbf{r}_\tau} \hat{\tau}^x_{\mathbf{r}_\tau+2\mathbf{b}},
\end{align}
with
\begin{align}
  \hat{Q}^{\tau}_{\mathbf{r}_\tau}
  = \prod_{\tilde{n}^{}_{\mathrm{CM}}({\mathbf{r}_\tau'}) < \tilde{n}^{}_{\mathrm{CM}}({\mathbf{r}_\tau})} \hat{\tau}^z_{\mathbf{r}_\tau'}.
\end{align}
The commutations relations for $\mu$ spins can be verified in a similar way.

\begin{figure*}
  \centering
  \includegraphics[width=0.75\hsize,bb=0 0 910 659]{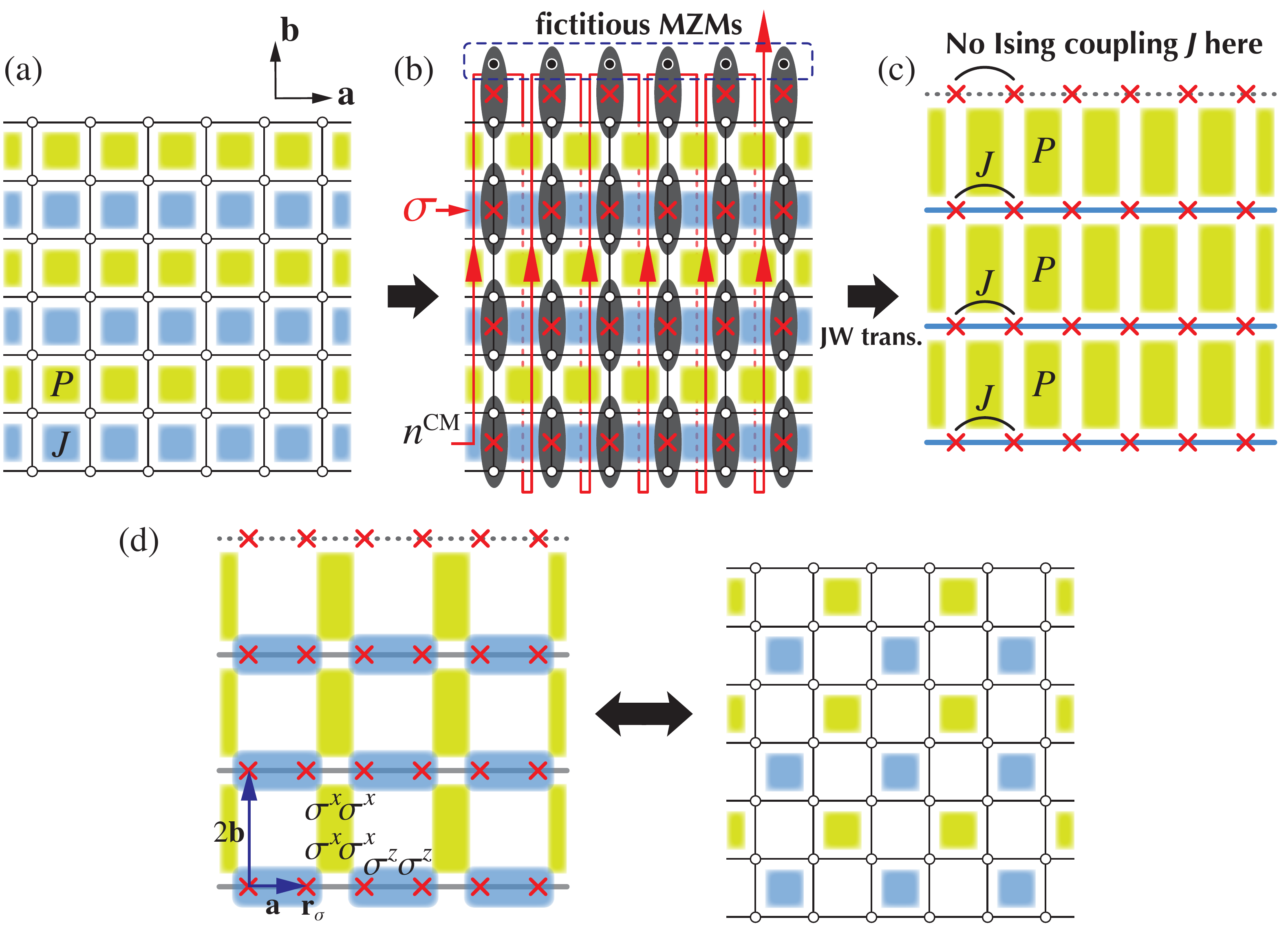}  
  \caption{%
    \label{fig:JW}
    (a) Lattice of Majorana fermions with an odd number of rows in the $\mathbf{b}$ direction.
    (b) JW transformation.
    (c) Lattice of $\sigma$ spins.
    (d) Illustration of the components of the order parameter $\hat{D}_\sigma$ [Eq.~\eqref{eq:SM:D}] and their relations with the plaquette interaction in the Majorana fermion system.
  }
\end{figure*}

\section{Trick of fictitious Majorana zero mode}

Our derivation of $\Hsigma$ in the main text assumes an even number of rows in the square lattice of Majorana fermions. Because of the open boundary in the $\mathbf{b}$ direction, this means there are an odd number of plaquettes in this direction. However, since the low $T$ phase spontaneously induces a modulation of energy density associated with the plaquette interaction $g$, a particular type of the modulation is favored in such a setup, even if there are an even number of plaquettes in the $\mathbf{a}$ direction. To avoid a resulting finite-size effect in the QMC simulation, we consider the lattice with an even number of plaquettes in the $\mathbf{b}$ direction, hence, an odd number of rows of Majorana fermions [Fig.~\ref{fig:JW}(a)]. 

A faithful spin representation can be derived also in this case. To combine the Majorana fermions in pairs properly, we add an additional row of \emph{fictitious} MZMs as shown in Fig.~\ref{fig:JW}(b). These fictitious modes are neither coupled to the rest of the system nor hybridizing within themselves. Therefore, they simply contribute to a constant to the free energy. We then apply the JW transformation in the same way as described in the main text. The only difference in the final form of $\Hsigma$ is that the horizontal $J$ coupling for $\sigma$ spins in the top row of the lattice is absent in the present case [Fig.~\ref{fig:JW}(c)], because it would correspond to an interaction involving the fictitious modes, which do not exist.
Finally, Fig.~\ref{fig:JW}(d) illustrates each component of the order parameter of the Majorana stripe state,
\begin{align}
  \hat{D}_\sigma = \frac{2}{L_a(L_b-1)} \sum_{\mathbf{r}_\sigma \in \textrm{even (or odd)}}
  \hat{D}_\sigma({\mathbf{r}_\sigma}),
  \label{eq:SM:D}
\end{align}
where $L_a$ and $L_b$ are the size of the lattice of $\sigma$ spins in the horizontal and vertical directions, respectively, and $\hat{D}_\sigma({\mathbf{r}_\sigma})$ is given in Eq.~(8) in the main text.

\section{Mean field theory}

Here we present the details of our zero-temperature MF calculation of the Majorana stripe phase. To consider the more general situations, we introduce two coupling constants $g$ and $g'$ for plaquettes on the even and odd-numbered rows, respectively, of the square lattice (they correspond to $P$ and $J$, respectively, in the main text). The difference $\Delta g \equiv g - g'$ serves as a symmetry-breaking field. As a result of this explicit breaking of translation symmetry along the vertical direction, the unit cell is doubled:
\begin{align}
  \hat{{H}}
  &=
  g \sum_{\mathbf{r}_\sigma}
  \hat{\gamma}^{}_{\mathbf{r}_\sigma,1}
  \hat{\gamma}^{}_{\mathbf{r}_\sigma - 2\mathbf{b},2}
  \hat{\gamma}^{}_{\mathbf{r}_\sigma + \mathbf{a},1}
  \hat{\gamma}^{}_{\mathbf{r}_\sigma + \mathbf{a} - 2\mathbf{b},2}
  \notag\\
  &+ g' \sum_{\mathbf{r}_\sigma}
  \hat{\gamma}^{}_{\mathbf{r}_\sigma,2}
  \hat{\gamma}^{}_{\mathbf{r}_\sigma,1}
  \hat{\gamma}^{}_{\mathbf{r}_\sigma + \mathbf{a},2}
  \hat{\gamma}^{}_{\mathbf{r}_\sigma + \mathbf{a},1}
  \notag\\
  &+
  i t \sum_{\mathbf{r}_\sigma}
  \left(
  \hat{\gamma}^{}_{\mathbf{r}_\sigma,2} \hat{\gamma}^{}_{\mathbf{r}_\sigma,1}
  +
  \hat{\gamma}^{}_{\mathbf{r}_\sigma,1} \hat{\gamma}^{}_{\mathbf{r}_\sigma - 2\mathbf{b},2}
  \right)
  \notag\\
  &+
  i t \sum_{\mathbf{r}_\sigma}
  \left(
  \hat{\gamma}^{}_{\mathbf{r}_\sigma,2} \hat{\gamma}^{}_{\mathbf{r}_\sigma + \mathbf{a},2}
  -
  \hat{\gamma}^{}_{\mathbf{r}_\sigma,1} \hat{\gamma}^{}_{\mathbf{r}_\sigma + \mathbf{a},1}
  \right),
\end{align}
where the previous notation of $\hat{\gamma}^{}_{\mathbf{r}_\sigma,s}$ ($s=1,2$) is used [see Fig.~1(b) in the main text]. We also note that $t \ne 0$ lifts the 1D gauge-like symmetries discussed in the main text.

We first consider the most general MF decouplings of the quartic terms assuming no further breaking of the translation symmetry. Direct numerical calculation nonetheless shows that the diagonal term such as
$\langle{ \hat{\gamma}^{}_{\mathbf{r}_\sigma,1} \hat{\gamma}^{}_{\mathbf{r}_\sigma + \mathbf{a},2} }\rangle$
vanishes in the self-consistent solution. We thus consider the following nonzero MF averages:
\begin{subequations}
\begin{align}
  \Delta &= i \langle{ \hat{\gamma}^{}_{\mathbf{r}_\sigma,2} \hat{\gamma}^{}_{\mathbf{r}_\sigma,1} }\rangle,
  \label{eq:MF1}
  \\
  \Delta'&= i \langle{ \hat{\gamma}^{}_{\mathbf{r}_\sigma,1} \hat{\gamma}^{}_{\mathbf{r}_\sigma - 2\mathbf{b},2} }\rangle,
  \label{eq:MF2}
  \\
  \delta &= i \langle{ \hat{\gamma}^{}_{\mathbf{r}_\sigma,2} \hat{\gamma}^{}_{\mathbf{r}_\sigma + \mathbf{a},2} }\rangle,
  \label{eq:MF3}
  \\
  \delta' &= i \langle{ \hat{\gamma}^{}_{\mathbf{r}_\sigma,1} \hat{\gamma}^{}_{\mathbf{r}_\sigma + \mathbf{a},1} }\rangle.
  \label{eq:MF4}
\end{align}
\end{subequations}
The resultant MF Hamiltonian is
\begin{align}
  \hat{H}_\mathrm{MF}
  &=
  -2 i g \Delta \sum_{\mathbf{r}_\sigma} \hat{\gamma}^{\;}_{\mathbf{r}_\sigma, 2} \hat{\gamma}^{\;}_{\mathbf{r}_\sigma,1}
  - 2 i g' \Delta' \sum_{\mathbf{r}_\sigma} \hat{\gamma}^{\;}_{\mathbf{r}_\sigma, 1} \hat{\gamma}^{\;}_{\mathbf{r}_\sigma - 2\mathbf{b},2}
  \notag\\
  &+ i (g+g') \delta \sum_{\mathbf{r}_\sigma} \hat{\gamma}^{\;}_{\mathbf{r}_\sigma,1} \hat{\gamma}^{\;}_{\mathbf{r}_\sigma + \mathbf{a}, 1}
  + i (g+g') \delta' \sum_{\mathbf{r}_\sigma} \hat{\gamma}^{\;}_{\mathbf{r}_\sigma,2} \hat{\gamma}^{\;}_{\mathbf{r}_\sigma + \mathbf{a}, 2} 
  \notag\\
  &+
  i t \sum_{\mathbf{r}_\sigma}
  \left(
  \hat{\gamma}^{}_{\mathbf{r}_\sigma,2} \hat{\gamma}^{}_{\mathbf{r}_\sigma,1}
  +
  \hat{\gamma}^{}_{\mathbf{r}_\sigma,1} \hat{\gamma}^{}_{\mathbf{r}_\sigma - 2\mathbf{b},2}
  \right)
  \notag\\
  &+
  i t \sum_{\mathbf{r}_\sigma}
  \left(
  \hat{\gamma}^{}_{\mathbf{r}_\sigma,2} \hat{\gamma}^{}_{\mathbf{r}_\sigma + \mathbf{a},2}
  -
  \hat{\gamma}^{}_{\mathbf{r}_\sigma,1} \hat{\gamma}^{}_{\mathbf{r}_\sigma + \mathbf{a},1}
  \right) + \text{const.},
\end{align}
where the constant term is $N [g \Delta^2 + g' \Delta'^2 - (g + g') \delta \delta']$. The MF spectrum is obtained by solving $\hat{H}_\mathrm{MF}$ self-consistently with Eqs.~\eqref{eq:MF1}--\eqref{eq:MF4}.
We find the relation $\delta' = -\delta$ from our numerical solutions.

Figure \ref{fig:mf} shows the MF parameters as a function of the ratio $g' /g$ for two different hybridization constant $t /g = 0.2$ and $t/g =0.8$. The stripe order parameter $D$ defined in the main text is given by $D = \Delta - \Delta'$. For the small hopping $t/g = 0.2$ [Fig.~\ref{fig:mf}(a)], the stripe order $D$ remains finite as the system approaches the symmetric limit $g' \to g$. This result indicates the existence of a zero-temperature first-order transition at $\Delta g = 0$, discussed in the main text. As the hopping increases, the discontinuity in $D$ at $\Delta g = 0$ is also reduced and eventually vanishes when $t > t_c \approx 0.65g$.

\begin{figure*}[]
\center
\includegraphics[width=0.7\hsize,bb=0 0 950 450]{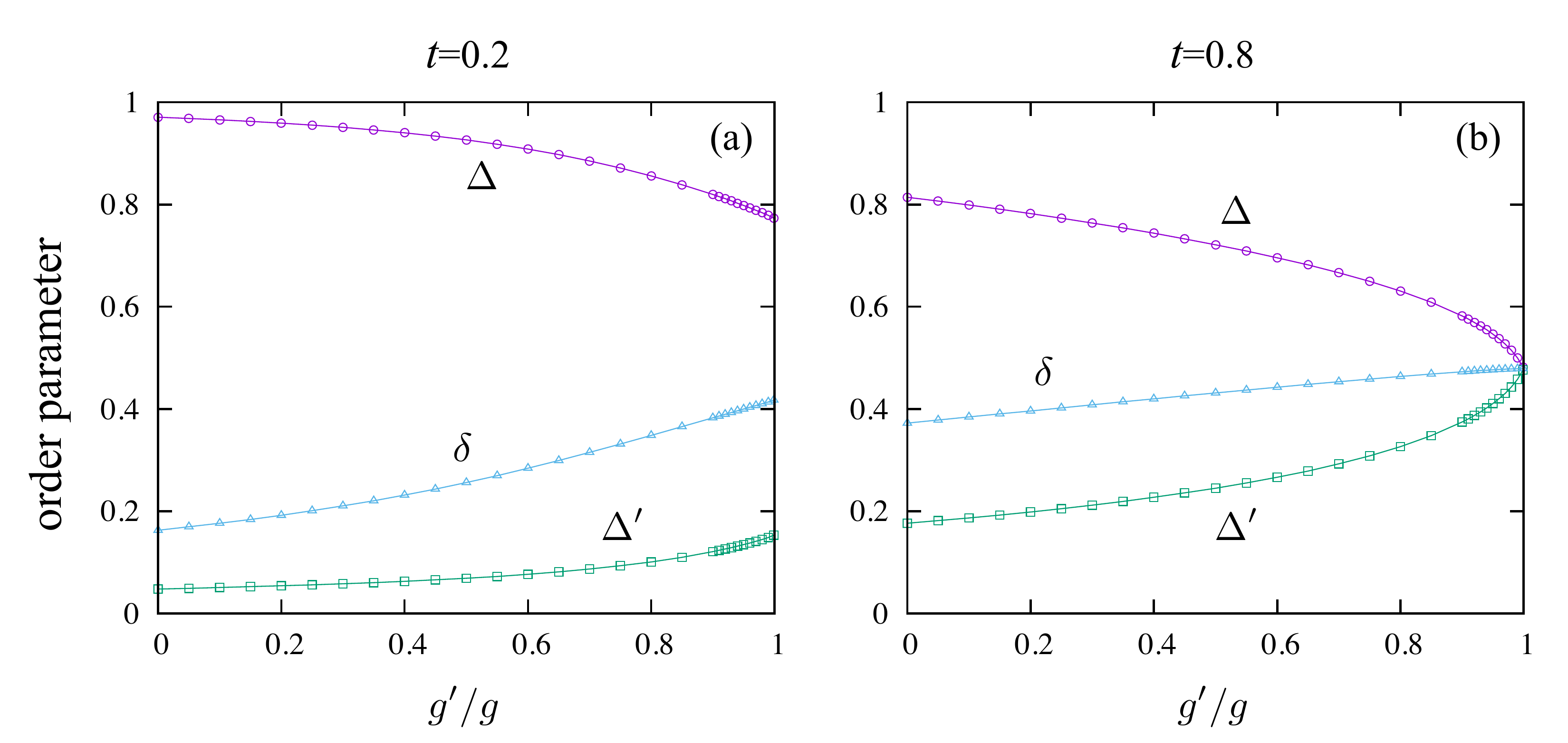}
\caption{%
MF order parameters $\Delta$, $\Delta'$, and $\delta$ as a function of the ratio $g'/g$ for (a) $t/g = 0.2$ and (b) $t/g = 0.8$.
\label{fig:mf}
}
\end{figure*}

We next perform the Fourier transform $\hat{\gamma}^{\;}_{\mathbf{r}_\sigma,s} = (N/2)^{-1/2} \sum_{\mathbf k} \hat{\gamma}_{\mathbf k, s} e^{i \mathbf k \cdot (\mathbf r + \mathbf d_s)}$, where $N/2$ is the number of unit cells and the basis vectors are $\mathbf d_{1, 2} = \pm \mathbf b/2$. The MF Hamiltonian can be expressed as $\hat{H}_{\rm MF} = \sum_{\mathbf k} \hat{\Psi}_{\mathbf k}^\dagger \mathpzc{H}_\mathrm{\,MF} \hat{\Psi}_{\mathbf k}$ with $\hat{\Psi}_{\mathbf k} \equiv (\hat{\gamma}^{\;}_{\mathbf{k},1},\, \hat{\gamma}^{\;}_{\mathbf{k},2})^\mathrm{T}$ and
  \begin{align}
    \mathpzc{H}_{\mathrm{\,MF}}
    = - [ t + (g + g') \delta] \sin k_x \, 
    \tau^z
    - (g \Delta + g' \Delta' - t) \sin k_y \, 
    \tau^x&
    \notag\\
    + (g \Delta - g' \Delta') \cos k_y \, 
    \tau^y&,
    \label{eq:H_MF}
  \end{align}
where $\tau^{x, y, z}$ are the Pauli matrices and we have used the relation $\delta' = -\delta$ we found numerically.
The MF band structure of Majorana fermions is shown in Fig.~\ref{fig:bandstructure} for various ratios of $g'/g$ with $t/g = 0.1$.
Expressing 
$\mathpzc{H}_\mathrm{\,MF}(\mathbf k) = \sum_{m = x, y, z} a_m(\mathbf k) \tau^m$, 
the eigenenergy of the MF Hamiltonian is given by $\varepsilon^{\pm}_{\mathbf k} = \pm \sqrt{a_x^2 + a_y^2 + a_z^2}$, and always appears in $\pm$ pairs. The energy gap for $t < t_c$ corresponds to the minimum of $\varepsilon_{\rm gap} = \min (\varepsilon^+_{\mathbf k} - \varepsilon^-_{\mathbf k})$ occurs at $\mathbf k = (0,0)$, $(\pi, 0)$, $(0, \pi)$, and $(\pi, \pi)$. In the symmetric point $g = g'$, the spectral gap is related to the stripe order parameter
\begin{align}
   \varepsilon_{\rm gap} = 2  g (\Delta - \Delta') =  g \Delta D.
\end{align}              
The energy gap as a function of the hybridization is shown in Fig.~5 of the main text. Importantly, the closing of the gap $\varepsilon_{\rm gap} = 0$ for $t > t_c$ gives rise to a critical state with low-energy gapless Majorana fermions; see Fig.~\ref{fig:bandstructure}.

\begin{figure}[]
   \center
   \if\arXiv1
   \includegraphics[width=\columnwidth,bb=0 0 1024 768]{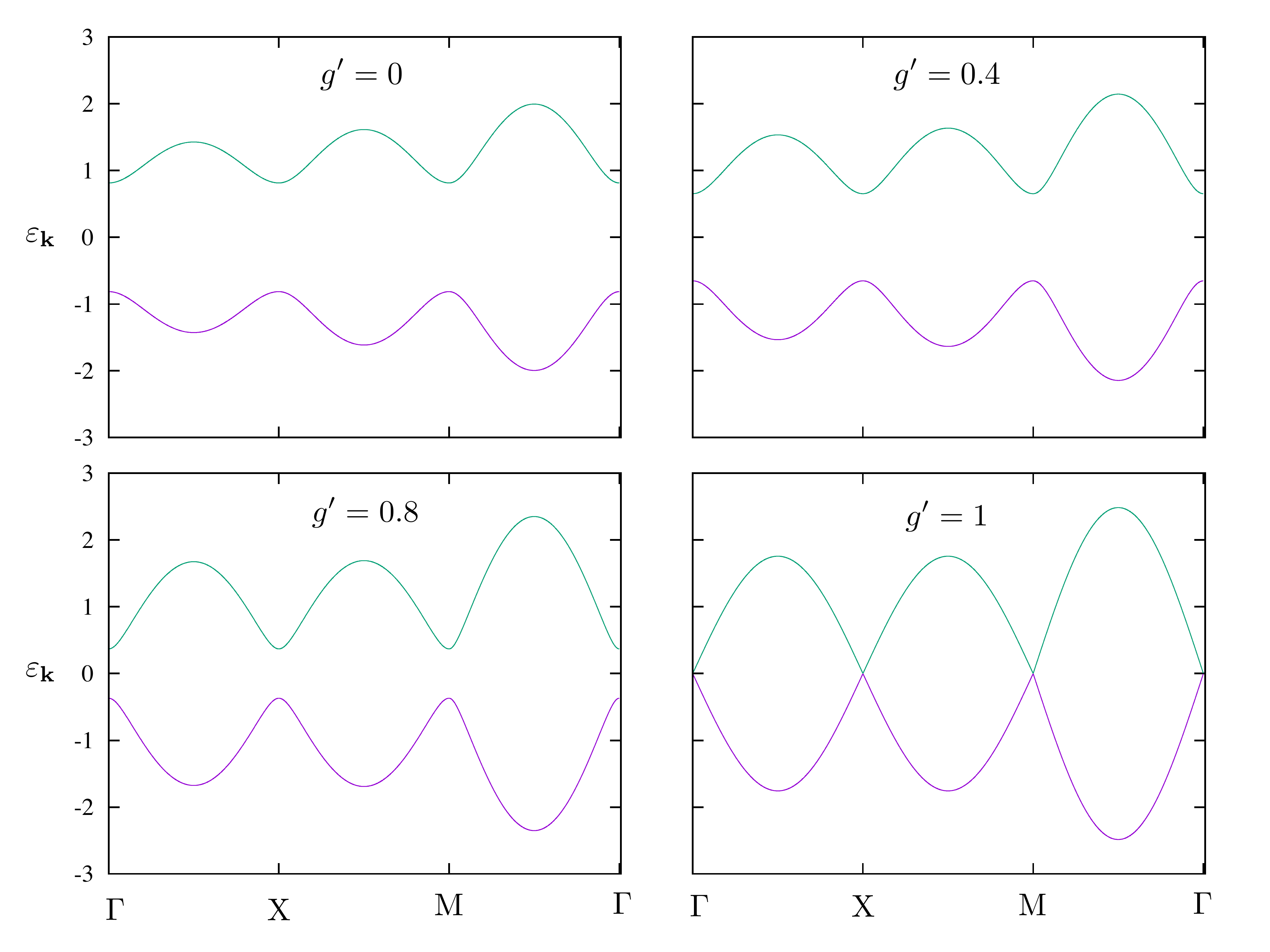}
   \fi
   \if\arXiv0
   \includegraphics[width=0.75\columnwidth,bb=0 0 1024 768]{FIG_S4_spectrum.pdf}
   \fi
   \caption{%
   \label{fig:bandstructure}
   MF quasi-particle band structure for $t/g = 0.8$.
   }%
\end{figure}

\begin{figure}
\center
\if\arXiv1
\includegraphics[width=0.7\hsize,bb=0 0 600 600]{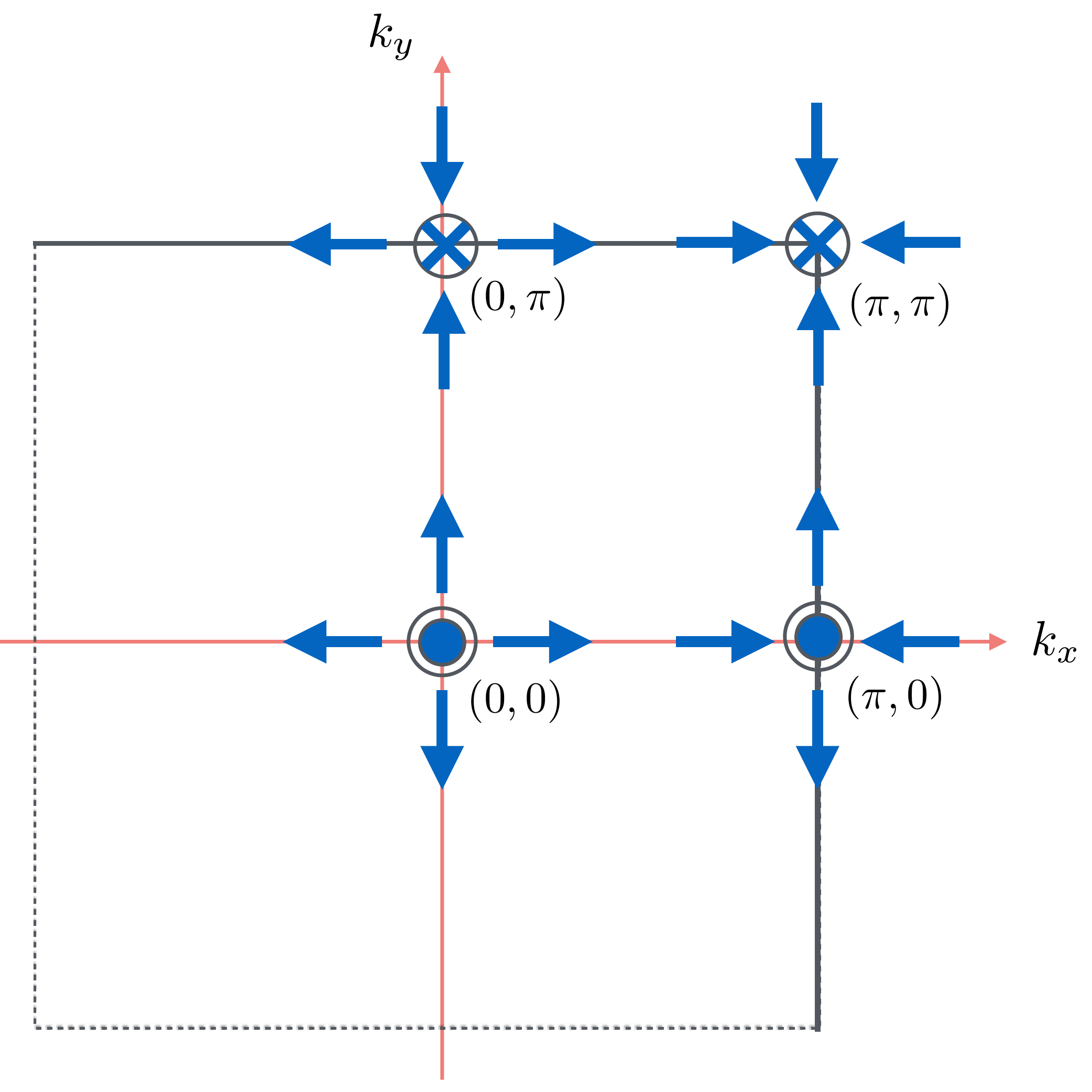}
\fi
\if\arXiv0
\includegraphics[width=0.4\hsize,bb=0 0 600 600]{FIG_S5_BZ.pdf}
\fi
\caption{%
\label{fig:BZ}  
The vector field in the vicinity of the four gapless Majorana nodes: $\mathbf m \approx (k_x, k_y, C_z)$ around $(0, 0)$, $\mathbf m \approx (-k_x, k_y, C_z)$ around $(\pi, 0)$, $\mathbf m \approx (k_x, -k_y, -C_z)$ around $(0, \pi)$, and $\mathbf m \approx (-k_x, -k_y, -C_z)$ around $(\pi, \pi)$.
}
\end{figure}

Starting from large hybridization for $t > t_c$ (focusing on the symmetric case $g = g'$), the Majorana stripe order occurs through a quantum phase transition that gaps out the Majorana nodal points upon reducing the hopping $t$. A natural question is whether the gapped MF band structure is topologically nontrivial. To answer this question, we compute the Chern number of the bands explicitly for $t < t_c$. We first define a unit-length vector $\mathbf m(\mathbf k) = [a_z(\mathbf k), a_x(\mathbf k), a_y(\mathbf k)] / \varepsilon^+_{\mathbf k}$. The spectral Chern number is given by
\begin{align}
        \mathcal{C} = \frac{1}{4\pi} \iint \mathbf m \cdot \left(\frac{\partial \mathbf m}{\partial k_x} \times \frac{\partial \mathbf m}{\partial k_y}  \right) dk_x dk_y = \mbox{integer}.
\end{align}
For simplicity, we introduce coefficients $C_{x, y, z}$ such that $\mathbf m(\mathbf k) = [C_x \sin k_x, C_y \sin k_y, C_z \cos k_y]/\varepsilon^+_{\mathbf k}$; these coefficients can be easily obtained by comparing with Eq.~(\ref{eq:H_MF}).
In our case, the integrand $\mathbf m \cdot \partial_{k_x} \mathbf m \times \partial_{k_y} \mathbf m$ evaluates to $C_x C_y C_z \, \cos k_x  / (\varepsilon^+_{\mathbf k})^3$. Consequently, the Chern number is zero, indicating a topologically trivial gapped phase. This result can also be understood by noting that the $\mathbf m$-vector maps the Brillouin zone to a unit sphere, and the Chern number is simply the winding number of this mapping. In the gapped phase (say $C_z = g D> 0$), the neighborhood around the original Majorana nodes $(0, 0)$ and $(\pi, 0)$ is mapped to the north hemisphere, while that around the other two nodes is mapped to the south hemisphere; see Fig.~\ref{fig:BZ}. Within each hemisphere, the winding number is determined by the in-plane vorticity of the two nodes. In our case, the two Majorana nodes within the same hemisphere have opposite vorticity $\pm 1$, hence the net winding number is zero.

\fi

\end{document}